# Magnetization reversal in Kagome artificial spin ice studied by first-order reversal curves

L. Sun[1], C. Zhou[1], J. H. Liang[1], T. Xing[2], N. Lei[2], P. Murray[3], Kai Liu[3], C. Won[4], and Y. Z. Wu[1,5*]

1 Department of Physics, State Key Laboratory of Surface Physics, and Advanced Materials Laboratory, Fudan University, Shanghai 200433, China

2 Fert Beijing Institute, School of Electronic Information Engineering, Beihang University, Beijing 100191, China

3 Department of Physics, University of California, Davis, California 95616, USA

4 Department of Physics, Kyung Hee University, Seoul 130-701, Republic of Korea

5 Collaborative Innovation Center of Advanced Microstructures, Nanjing 210093, China

Abstract

Magnetization reversal of interconnected Kagome artificial spin ice was studied by the first-order reversal curve (FORC) technique based on the magneto-optical Kerr effect and magnetoresistance measurements. The magnetization reversal exhibits a distinct six-fold symmetry with the external field orientation. When the field is parallel to one of the nano-bar branches, the domain nucleation/propagation and annihilation processes sensitively depend on the field cycling history and the maximum field applied. When the field is nearly perpendicular to one of the branches, the FORC measurement reveals the magnetic interaction between the Dirac strings and orthogonal branches during the magnetization reversal process. Our results demonstrate that the FORC approach provides a comprehensive framework for understanding the magnetic interaction in the magnetization reversal processes of spin-frustrated systems.

I. Introduction

An artificial spin ice system is made by patterning magnetic materials into nanoarrays of single domain ferromagnetic islands or connected nanowires, in which the strong shape anisotropy determines the magnetization orientation and results in Ising macrospin behavior. The closely arranged nano-bars or connected nanowires give rise to spin frustration, which contains energetically equivalent micromagnetic states [1-6]. The magnetic frustration can be modulated by patterning the nanomagnetic arrays into different dimensions and arrangements, and thus the spin frustration can be directly studied by real space imaging techniques [2,3, 7 - 10 ]. The artificial structures provide insight into fundamental understanding of magnetic frustration, especially in the degenerate states and charge-ordered states [2, 5, 10- 14 ]. Although artificial spin ice was designed originally to study thermodynamics of isolated nano-bars, recent works have also focused on the field-driven dynamics in interconnected spin ice structures [7-9]. In the spin ice structures, the magnetization reversal is usually controlled by the ice rule that governs the number of magnetizations pointing into and out of each vertex to minimize the local magnetostatic energy [8,13]. The magnetization reversal process in the spin ice structures has been systematically studied by real-space imaging techniques [7−9,15-18], magnetic hysteresis loop measurements using the magneto-optic Kerr effect [19,20], and magnetoresistance (MR) measurements [16,21−23]. The correlation between the magnetization switching and the associated MR change during the reversal process was carefully investigated recently [23].

It is believed that interconnected artificial spin ice reverses through the nucleation and propagation of magnetic domain walls (DWs). The magnetic DWs propagate through an interconnection and trigger spins in a neighboring bar to flip, which leads to chains of magnetization reversal. Such chains of overturned magnetic moment are called "Dirac strings", along which the spin ice rule is maintained except for the two ends [7-9]. The magnetization reversal in interconnected Kagome artificial spin ice also has a strong angular dependence [20]. Artificial spin ice is a strongly correlated system with high spin disorder, which contains complex interactions among the branches and interconnections during the magnetization reversal. The complex magnetic interactions are important to understand the magnetization reversal process. However, the magnetic interactions occurring in the reversal process are difficult to probe by static real-space magnetic imaging or hysteresis loop measurements.

The first-order reversal curve (FORC) technique has been widely used for the magnetization characterization in magnetic nanostructures [24–28]. It provides information about irreversible magnetic switching [29,30], magnetic interactions [31–33], distributions of magnetic characteristics [34,35], and magnetic phase separation [36,37], which are not easily accessible in conventional hysteresis loop investigations. While most FORC studies have been based on magnetometry measurements (M-FORC), recently the FORC methodology has also been extended to transport measurements such as magnetoresistance curves (MR-FORC) to probe spin disorder [38], and temperature-dependent resistivity measurement to investigate first order phase transitions [39,40].

In this paper, we have investigated the magnetization reversal processes in Kagome artificial spin ice by both M-FORC and MR-FORC measurements to gain a comprehensive understanding of the reversal mechanism. When the field is parallel to one of the branches, the domain nucleation/propagation and annihilation processes sensitively depend on the magnetic history and the maximum reversal field applied, which is difficult to distinguish from conventional hysteresis loop measurements alone. With the field nearly perpendicular to one of the branches, the FORC measurements exhibit rich features that clearly reveal the magnetic interactions between the Dirac strings and the orthogonal branches during the magnetization reversal. Moreover, the MR-FORCs expose many features which are not observable in M-FORCs. It shows that the dynamics of spin-frustrated systems can be comprehensively understood when the information from the MR-FORC and the M-FORC are combined together.

II. Experiment

Figure 1(a) shows a typical interconnected Kagome artificial spin ice nanostructure studied in this paper. The dimensions of the bars are 1μm in length and 150nm in width with a thickness of 20nm. The structures were patterned by e-beam lithography, followed by e-beam evaporation and lift-off of 20nm $Ni_{80}Fe_{20}$ onto a Si/SiO$_2$ substrate. A 2 nm thick SiO$_2$ capping layer was grown to prevent oxidation. The electrode pads were fabricated by the same lithography and lift-off process on the Cr/Au electrodes grown by magnetron sputtering.

Magnetic properties of Kagome artificial spin ice were investigated by a commercial focused magneto-optical Kerr effect (MOKE) Microscope (NanoMOKE_3) at room temperature with a 660nm diode laser. The hysteresis loops were measured in the longitudinal MOKE geometry, and each hysteresis loop was averaged over 100 cycles to enhance the signal-to-noise ratio. The sample can be rotated in the sample plane with an azimuthal angle $\theta$ between one branch and the field direction, as shown in Fig. 1(a). The size of the laser spot is ∼30 $\mu$m, which is smaller than that of patterned artificial Kagome spin ice structure (∼100 $\mu$m).

Figure 1(b) shows the typical sample structure for transport measurements. The sample size for the transport measurement is 40 $\mu$m×20 $\mu$m. The longitudinal magnetoresistance of artificial spin ice structures was measured with the standard four-probe method. The electric current was injected from the two large electrodes and the voltage was measured by the two smaller pads at one side. A quadrupole magnet was used to produce a magnetic field with any desirable direction in the sample plane. The resistance was measured with a standard lock-in technique with the modulation frequency of ∼1117 Hz and the applied AC current of 100 $\mu$A.

III. Results and discussion

Figures 2(a)-(e) show typical hysteresis loops measured by MOKE with different field orientations. The hysteresis loops were measured with θ varied from 0° to 180° by every 5°. For the field along one of the branches in the spin ice structure, i.e. θ=0°, the loops show a single-step of the avalanche-like switching behavior [8,9,18]. When the field orientation is away from the directions of branches and close to the orthogonal direction of one of the three branches, the MOKE loops show a clear two-step feature. Such a two-step feature indicates that the magnetizations of three branches at each interconnection will reverse at two distinct fields: one group of the three branches first reverses under the smaller magnetic field, and the other group reverses at higher magnetic field. The change of magnetization at the lower switching field $H_1$ is much larger than that at the higher switching field $H_2$. Fig. 3(a) shows that the switching fields $H_1$ and $H_2$ have an angular dependence with a clear 60°-rotational

symmetry originated from the symmetry of the Kagome lattice geometry. The lower switching field $H_1$ is almost independent of the field orientation, while the higher switching field $H_2$ has the maximum when the field is perpendicular to one of the three branches. When the field orientation is close to one of the branches, e.g., $-10° < \theta < +10°$, only the lower switching fields $H_1$ are plotted in Fig. 3. In this angle range, the two switching fields $H_2$ and $H_1$ are too close to each other to separate. The remanent Kerr signal is smaller than the saturation Kerr signal for all field orientations, as the magnetization in each bar should be parallel to its orientation at the remanent state due to the shape anisotropy.

Comparing with the hysteresis loop measurements, the MR curves exhibit more complex behavior, as shown in Fig. 2(f)-(j). Regardless of the field direction, the resistance reaches the maximum near zero field due to the anisotropic MR (AMR) effect. In the presence of strong shape anisotropy, the remanent magnetization in each bar is along the current direction and it results in a maximum MR. Note that the MR signal still does not reach saturation at 2000 Oe. This non-saturated MR can be attributed to the fact that the magnetization in some branches still deviates from the external field direction even in the high field. However, the Kerr signals in Fig. 2(a)-(e) show very little change for H > 700 Oe. This difference indicates that the MR measurement may be more sensitive to non-collinear magnetic configurations than the usual magnetization measurements. During the field sweeping process, the resistance shows an obvious irreversible change due to the magnetization switching. Here, we focus on the ascending-field sweeps that are shown in red in Fig. 2. For $\theta = 0°$ and 15°, the resistance decreases with increasing positive field, with a sudden drop at $H_1$, and then conforms onto the same MR curve as measured in the descending-field sweep, consistent with the single-step switching behavior in the hysteresis loops. However, for $\theta = 35°$, 85° and 90°, after the first irreversible switching at $H_1$, the resistance in the ascending-field sweep still has different value compared with the descending-field sweep before they merge together at a higher field $H_2$. Fig. 3(b) shows the angular dependence of $H_1$ and $H_2$ obtained from MR curves as a function of field orientation angles, which also reveals the 6-fold symmetry. For $\theta = 90°$ in Fig. 2(j), $H_2$ is hard to define due to the strong MR hysteresis. The angular dependent switching fields measured by the MR measurements are very similar to those in Fig. 3(a) measured by the hysteresis loops, and the slight differences may be attributed to the different sample preparations in the two measurements.

To understand the spin-correlation and interactions during the magnetization reversal process in detail, we further performed the FORC analysis based on both MOKE and MR measurements. The FORCs were generally measured in the following process [41]: After positive saturation, the magnetization M is measured starting from a negative reversal field $H_r$ to positive saturation, tracing out one FORC. A family of FORCs are measured at different $H_r$. The FORC distribution is then defined by a mixed second-order derivative [38]:

$$\rho(H_r,H) \equiv -\frac{1}{2}\frac{d^2 A(H_r,H)}{dH_r dH} \quad (1)$$

where A corresponds to M or MR. The second order derivative eliminates the reversible magnetization process, thus a plot of the FORC distribution $\rho(H_r,H)$ can be created to probe details of the irreversible magnetization reversal and magnetic interaction [29,34]. The FORC distribution can be plotted on the ($H_C$, $H_B$) plane defined by local coercivity $H_C=(H-H_r)/2$ and interaction field $H_B=(H+H_r)/2$ [31]. We have applied a standard smoothing and interpolation process for calculating accurate FORC distribution, with the detailed method described in Ref. 25. The MOKE and MR measurements showed that the magnetization switching are very different for the field parallel or perpendicular to one of the branches in the spin ice system. Since the magnetization in the horizontal bar gradually rotates towards the field at $\theta = 90°$, our discussion will mainly focus on measurements with the field angle of $0°$ and $85°$, where either the clear one-step or two-step switching is observed in both MOKE and AMR measurements.

A set of representative M-FORCs and the corresponding M-FORC distribution in the (H,$H_r$) coordinate system with $\theta = 0°$ are shown in Figs. 4(a) and (b), respectively. The major loop in Fig. 4(a) exhibits a sharp single-step switching over a small field range, which corresponds to an avalanche-like magnetization reversal throughout the system [8,9]. When the sample arrays are partially reversed with smaller negative $H_r$, these magnetic arrays can switch back to the saturation condition under increasing H. The corresponding M-FORC distribution shows a "left-bending boomerang" feature [33], which consists of a horizontal ridge for $H_r$>-330 Oe and a valley–peak pair for more negative $H_r$. Such a "left-bending boomerang" feature in the corresponding M-FORC distribution is typically associated with systems that exhibit domain nucleation and abrupt propagation [33], where domain growth is dominated by a strong exchange interaction [29]. Due to the presence of the exchange and dipolar interaction, the magnetization reversal in the spin-ice system usually takes place via DW nucleation from the sample edges, and propagate through the interconnections to form the Dirac strings [8, 9,18]. For the FORC curves with $H_r$ > -330 Oe, the onset of the up-switching field is almost independent of $H_r$, as indicated by the vertical dashed line in the inset in Fig. 4(a). This behavior corresponds to the horizontal ridge in the FORC distribution in Fig. 4(b). Further decrease of $H_r$ leads to the switching of residual moments, which not only requires a larger negative field to annihilate along the descending-field sweep, but also affects magnetization reversal in the subsequent ascending-field sweep, analogous to that seen in perpendicular Co/Pt multilayers [29]. For $H_r$ < -330 Oe, the onset of the up-switching field along each FORC gradually shifts with $H_r$, as indicated by the arrows in the blue curve with $H_r$ = -360 Oe and the red curve with $H_r$ = -330 Oe. This shift leads to the valley-peak pair feature in the FORC distribution. Therefore, these M-FORC results clearly indicate that the maximum reversal field the sample is exposed to not only determines how completely the residual domains are annihilated, but also affects the subsequent domain re-nucleation and propagation along the ascending-field sweep.

The magnetization reversal behavior at $\theta$ = 85° is markedly different from that at $\theta$ = 0°. Figures 5(a) and (b) show a set of representative M-FORCs and corresponding M-FORC distribution. The major hysteresis loop in Fig. 5(a) shows the two-step magnetization switching. The first step represents the magnetization switching of the up- and down-branch groups forming the zigzag Dirac strings [20], and the second step represents the switching of the horizontal branches perpendicular to the field. Owing to the strong shape anisotropy, a horizontal branch requires a higher switching field for a nearly perpendicular external field. The FORC distribution in Fig. 5(b) can be characterized by three main features, which are highlighted by the dashed circles with numbers. The first feature of M-FORC is a positive peak elliptically stretched along the $H_r$ axis for -370 Oe < $H_r$ < -280 Oe, and it is due to the magnetization up-switching process across the small range of H, as illustrated between the red and green curves in Fig. 5(a). For -520 Oe < $H_r$ < -370 Oe, the reversal curves are closely packed, thus the FORC distribution is almost zero. Further decrease of $H_r$ leads to the second feature with a negative/positive pair, in the M-FORC diagram. It arises from the mismatch between the ending point along the reversal curves, as indicated by the red and blue arrows in Fig. 5(a) inset. The third feature is a weak ridge for H > 600 Oe, which corresponds to the second up-switching event in the hysteresis loops.

The shape of the three features in the FORC diagram in Fig. 5 could be understood by the magnetization configurations shown in Fig. 6. Here, we only present a 4 × 4 matrix for illustration. At zero field after saturation in a strong positive field, the three branches at each interconnection show the "2-in-1-out" state or the "1-in-2-out" state due to the ice rule, as shown in Fig. 6(a). Fig. 6(b) illustrates the magnetic configuration at the switching field corresponding to the first FORC feature. The red arrows show the Dirac strings formed by the corresponding negative $H_r$ in the first feature, and those Dirac strings propagate back after a positive switching field H is applied. The up-switching behavior takes place over a small field range, while the nucleation field $H_r$ has a broader distribution. This is because the domain walls propagate avalanche-like after their nucleation. Moreover, in the first feature, the magnetization in the horizontal branches always points to the right, but it is tilted by H. The tilting component of the horizontal branches may induce an effective field to the Dirac strings through the exchange interaction and the dipolar interaction at the interconnection, thus promoting switching processes of the zigzag Dirac strings. This effective field can be identified in the M-FORC distribution in Fig. 5(b) since the center of the first feature is slightly away from the $H_c$ axis with the estimated bias field of ∼ 20 Oe, as illustrated by the dashed line.

Fig. 6(c) illustrates the magnetic configuration at the switching field in the second FORC feature. For -730 Oe <$H_r$< -520 Oe, the reversal field is large enough to switch down all the Dirac strings, as indicated by the red arrows. The partially reversed horizontal branches are indicated by the blue arrows. While the Dirac strings propagate back at the applied switching field, the magnetization in the horizontal branches tilts away from the bar direction, which provides an interaction field on the switching of the vertical Dirac strings. However, due to the switching field with $\theta = 85°$, the tilting angle for the reversed branches (blue arrows) is different from that for the unreversed ones (black arrows), resulting in the different interaction field to the reversal of the Dirac strings. The effective interaction field depends on the numbers of the reversed horizontal branches determined by $H_r$. From the switching fields in the FORC distribution at $H_r$=-730 Oe and $H_r$=-520 Oe, as indicated by the red arrows in the FORC diagram, we can estimate that the difference of the interaction fields from the reversed and unreversed horizontal magnetization is ~ 50 Oe.

The third feature in the M-FORC distribution in Fig. 5(b) is related to the magnetization switching of the horizontal branches for H > 400 Oe. This feature is parallel to the diagonal direction in the H-$H_r$ coordinate, i. e. along the $H_C$ axis in the $H_B$-$H_C$ coordinate, but slightly shifted along the negative $H_B$ axis. It is well known that the FORC distribution of non-interacting nanomagnetic arrays should spread along the $H_C$ axis [33,34]. Thus, the shifted feature away from the $H_C$ axis indicates that there exists an additional bias field during the up-switching of the horizontal branches. This bias field is estimated to be 50 Oe from the central $H_B$ field of the third feature. This bias field is much stronger than the dipolar interaction field between two parallel magnetic bars separated by 1μm, which is estimated to be 0.05 Oe. We attribute this bias field to the interaction between the Dirac strings and the horizontal branches. As indicated by Fig. 6(c), the magnetization in the Dirac string aligns nearly with the H direction, providing a bias field for the magnetization in the horizontal branches.

We applied the MR-FORC analysis on the spin-ice system as well to investigate more features of reversal process. As demonstrated in Ref. 30, FORC analysis based on the magnetoresistance curves has proven to be an effective method to study the microscopic magnetic configurations and the spin disorder. Fig. 7 (a) and (b) show a set of representative MR-FORCs and the corresponding MR-FORC distribution at $\theta = 0°$, respectively. As $H_r$ increases, the position of a local minimum in each MR curve shifts to more negative H, which is indicated by the arrows in Fig. 7(a). The corresponding MR-FORC distribution also shows a valley-peak pair but with a broader distribution along both H and $H_r$ axis than the M-FORC distribution in Fig. 4(b). This MR-FORC distribution contains similar features to the M-FORC distribution. The broad distribution can be attributed to the broad field span in the MR-FORCs, which may be attributed to the different sample preparation conditions for the M-FORC and MR-FORC measurements.

Figs. 7(c) and (d) show a set of representative MR-FORCs and the corresponding MR-FORC distribution for θ =85°. The three features in the M-FORC distribution also exist in the MR-FORC distribution in Fig. 5(b). The first feature in MR-FORC is slightly shifted from the $H_C$ axis, the second feature is tilted to higher H at the lower $H_r$ end, and the third feature is a valley-peak pair parallel to the $H_C$ axis. Since the magnetization reversal of the horizontal branches induces a clearer signal in the MR measurement than in MOKE measurement, it is expected that the third feature is more distinguishable than that in the M-FORC distributions. The MR change is not proportional to the magnetization change in the FORC measurement, so it is difficult to use features in the MR-FORC distribution to quantify the exchange field. Along with the three similar features as in the M-FORC distribution, the MR-FORC distribution also contains additional features, which are not observed in the M-FORC distribution. The fourth feature is a ridge, as highlighted in the solid oval for -370 Oe <$H_r$ <-290 Oe. It has symmetric distribution with H = 0 Oe. This feature can be attributed to the mismatch of slopes in the MR-FORCs in Fig. 7(c) with different $H_r$, during which the zigzag Dirac strings are partially formed. However, in the M-FORCs in Fig. 5(a), the hysteresis loops along the ascending-field sweep show the similar slope, resulting in a zero second-order derivative. This result indicates that, after formation of the Dirac strings, the microscopic magnetic configuration along the ascending-field sweeps depends on the number of the Dirac string, and the MR measurement is more sensitive to the change of the micromagnetic structure than the MOKE measurement. The fifth feature in Fig. 7(d) is highlighted in the solid rectangle as a valley. As shown in Fig. 6(c), the MR loops in MR-FORCs change the slope for -730 Oe < $H_r$ < -520 Oe, which results in the negative second-order derivatives. This feature also indicates that the microscopic spin structure can be strongly influenced by the reversal of the magnetization in the horizontal bars, thus the micromagnetic spin configurations and the degree of spin disorder are important in determining the MR, though the M-FORC measures a macroscopic magnetization.

IV. Conclusion

In summary, we have applied both M-FORC and MR-FORC techniques to investigate the magnetization reversal in connected Kagome artificial spin ice system. For the field parallel to one of the nano-bar branches, the magnetic history strongly influences the domain nucleation and annihilation process, which is hard to identify through hysteresis loop measurements alone. The maximum reversal field not only determines how completely the residual domains are annihilated, but also affects the subsequent domain re-nucleation and propagation process. For the field close to the orthogonal direction of one of the branches, the FORC distribution exhibits three features for the two-step magnetization reversal. These features correspond to the formation of zigzag Dirac strings, the horizontal branch switching, and the correlation between the deformation of the Dirac strings and the reversal of the horizontal branches. The FORC measurement clearly reveals the magnetic interaction between the Dirac strings and the horizontal branches in the magnetization reversal process. Our studies show that the M-FORC and MR-FORC measurements are complementary methods for understanding the irreversible magnetization reversal process for artificial spin ice systems. Moreover, our results also indicate that the MR-FORCs can sensitively reflect the microscopic magnetization configurations and the degree of total spin disorder, which are generally not observable by M-FORCs.

This project was supported by the National Key Basic Research Program of China (Grant No. 2015CB921401), the National Key Research and Development Program of China (Grant No. 2016YFA0300703), the National Science Foundation of China (Grants No. 11434003, No. 11474066, No. 11734006), and the Program of Shanghai Academic Research Leader (No. 17XD1400400). Work at UCD was supported by the US National Science Foundation DMR-1610060 (K.L.) and ECCS-1611424 (P.M.).

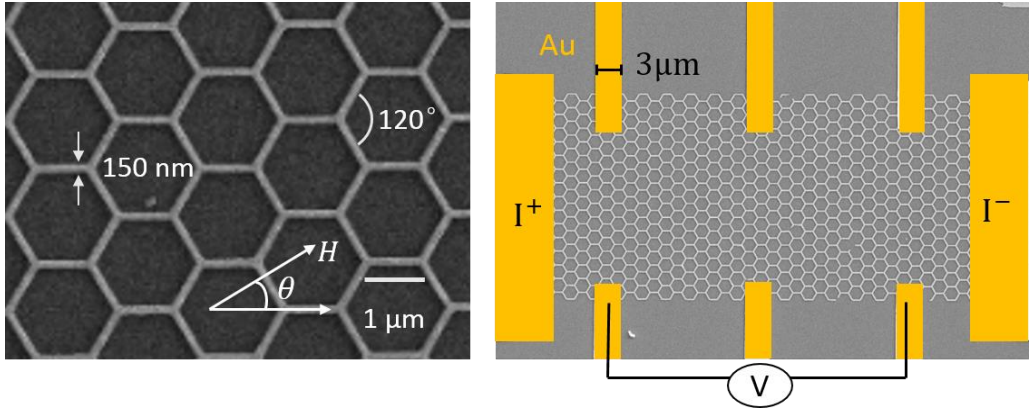

Fig. 1. (a) SEM image of an artificial spin ice structure with bar dimensions of 150 nm × 1 μm. External magnetic field H is applied at the angle θ relative to the structure. (b) SEM image of the transport measurement geometry with the Au/Cr electrodes artificially colorized.

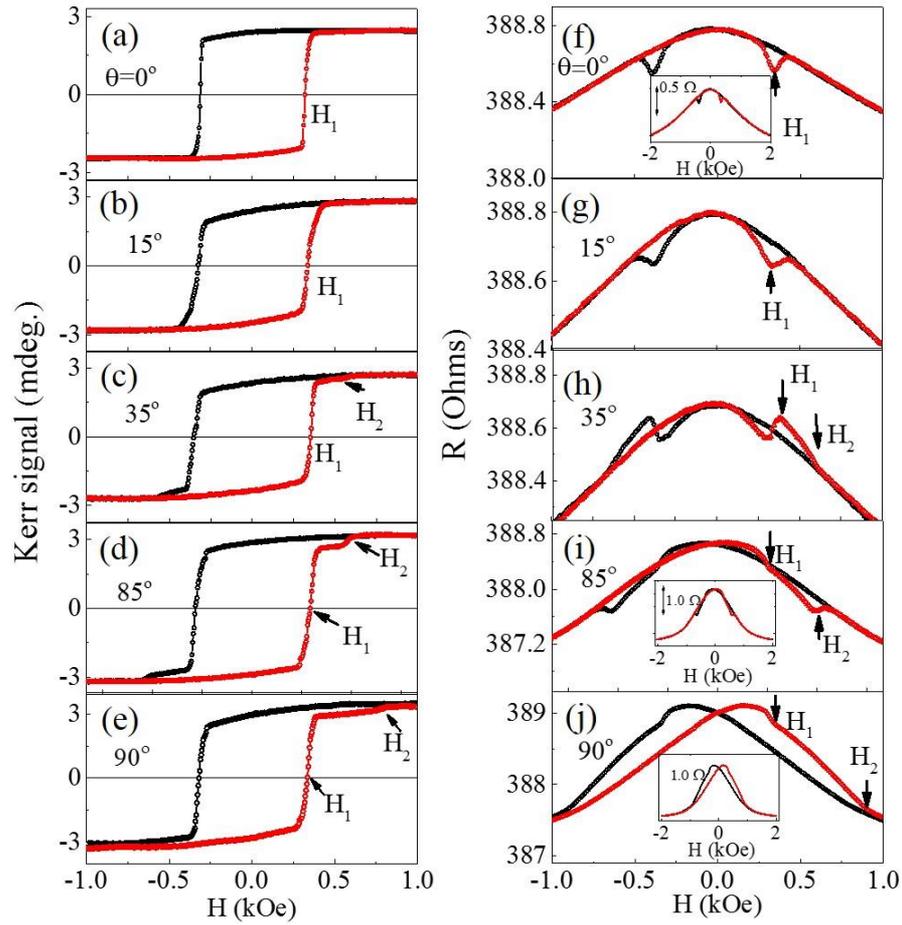

Fig. 2. Representative hysteresis loops (a-e) and magnetoresistance curves (f-i) of the spin ice sample with different external field angles. The insets in (f), (i) and (j) show the measured MR curves up to 2 kOe. $H_1$ and $H_2$ indicate the two distinct magnetization reversal fields, respectively.

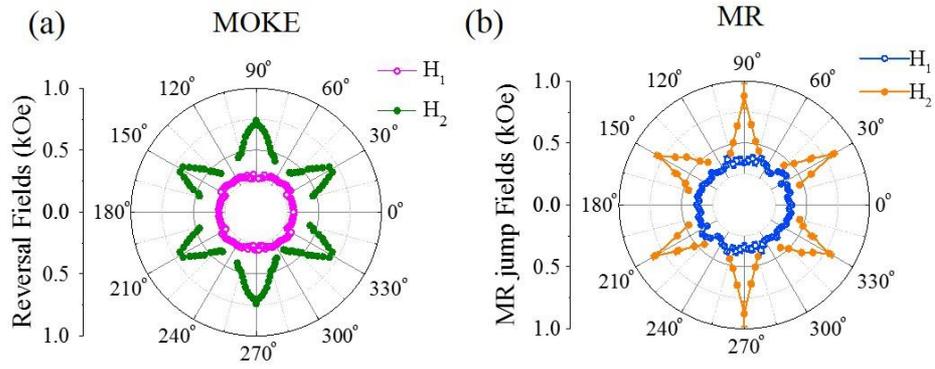

Fig. 3. Angular dependence of magnetization switching fields measured by (a) MOKE and (b) MR, as a function of the in-plane angle θ.

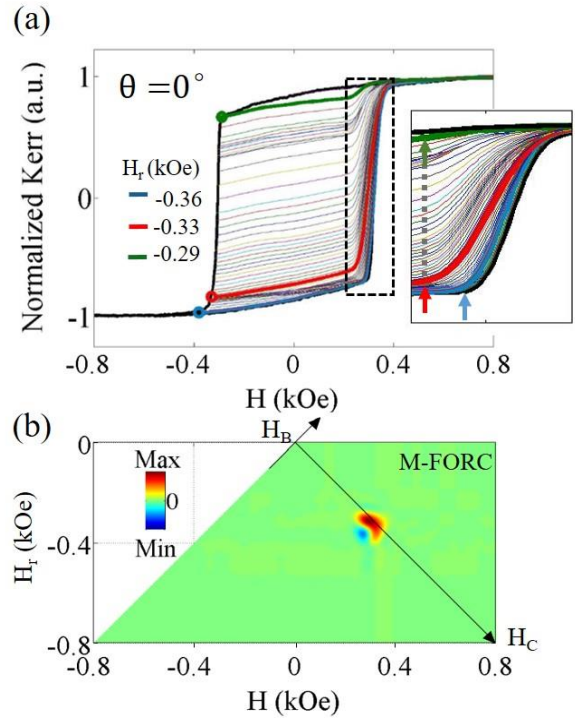

Fig. 4. (a) Representative curves of M-FORCs, with the highlighted curves starting from the indicated $H_r$ values. The switching behavior in the dashed rectangle region is magnified in the inset. The dashed line indicates that the onset switching field remains similar for the FORCs starting from -330 Oe < $H_r$ < -290 Oe. For $H_r$ < -330 Oe, the onset switching field of each reversing curve gradually shifts with $H_r$, as indicated by the arrows in the blue curve with $H_r$ = -360 Oe and the red curve with $H_r$ = -330 Oe. (b) The corresponding FORC distribution plotted both on the (H,$H_r$) and ($H_C$,$H_B$) coordinates.

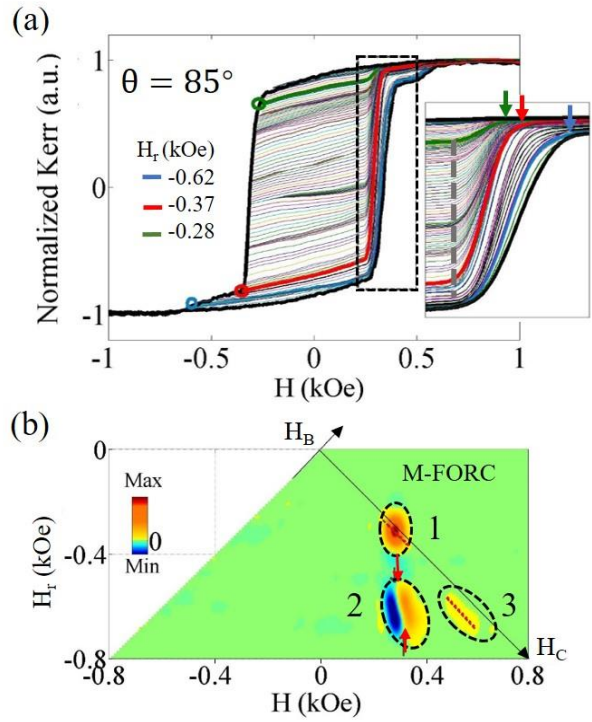

Fig. 5. (a) Representative curves of M-FORCs for θ=85°, with the highlighted curves starting from the indicated $H_r$ values. The switching behavior in the dashed region is zoomed in the inset. The dashed line in the inset indicates that the onset switching field remains similar for -370 Oe < $H_r$ < -280 Oe. The colored arrows indicate the mismatch between the ending points in the three highlighted curves. (b) The corresponding FORC distribution plotted both on the (H, $H_r$) and ($H_C$, $H_B$) coordinates. The ovals and numbers highlight the three features in the FORC distribution, with the arrows and the dashed red curves illustrating the shift of features discussed in the text.

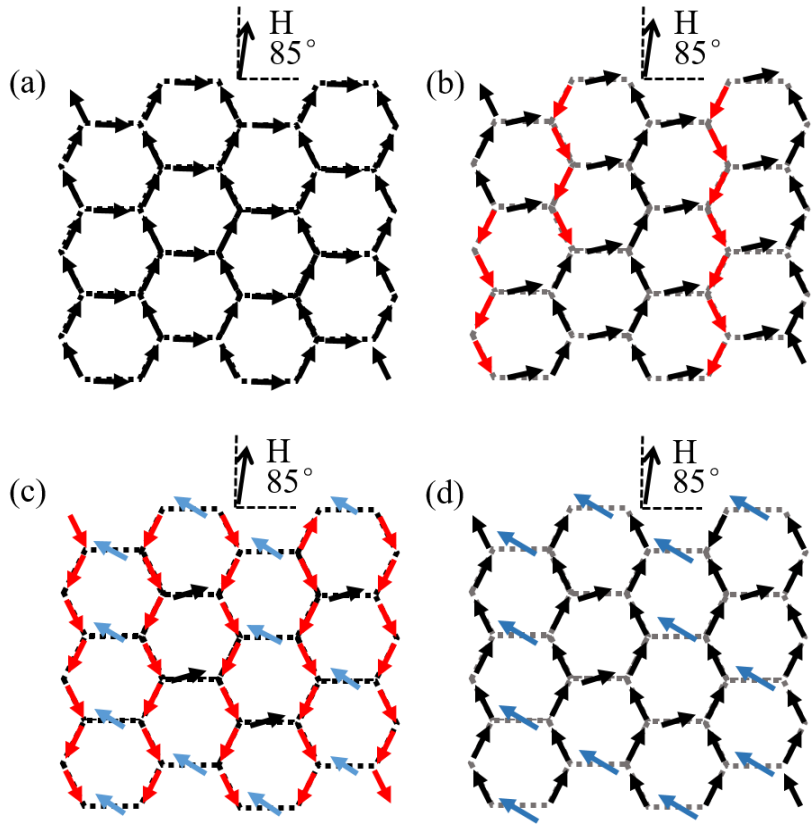

Fig. 6. Schematic illustration of magnetic configurations M (H, $H_r$) before switching at H for θ=85°, where $H_r$ corresponds to the starting field of each reversal curve, and H corresponds to the positive switching field. (a) A zero-field state relaxed from the positive field. (b)-(d) The states correspond to features 1-3 in Fig. 5(b), with (H, $H_r$) around (b) (350 Oe, -350 Oe), (c) (350 Oe, -600 Oe), and (d) (550 Oe, -600 Oe), and the colored arrows indicating the reversed branches for different $H_r$.

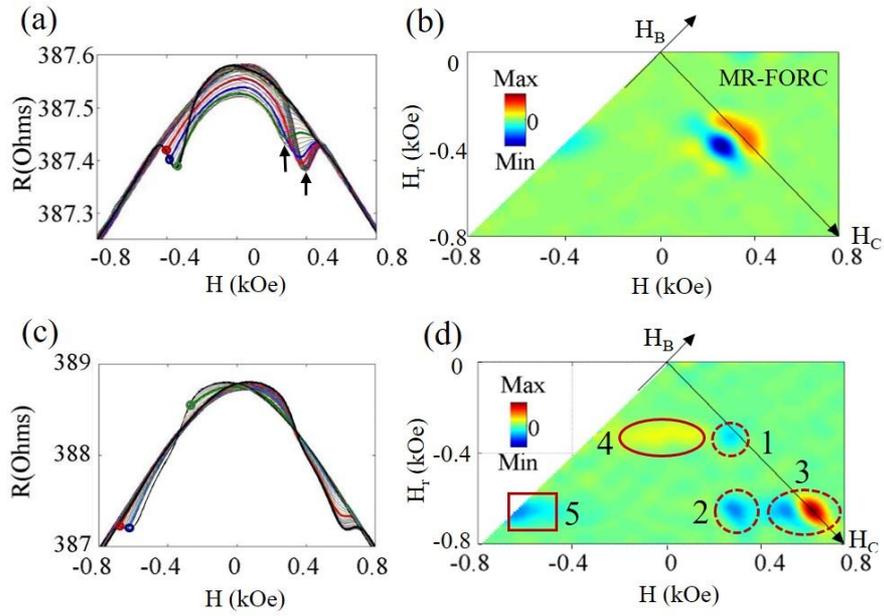

Fig. 7. MR-FORCs and corresponding MR-FORC distributions for $\theta = 0°$ ((a) and (b)) and $\theta=85°$ ((c) and (d)), respectively, with the highlighted curves starting from the indicated $H_r$ values. Arrows in (a) indicate the MR local minimum shifting with $H_r$, which corresponds to the left-bending feature in (b). The highlighted regions in (d) with the numbers indicate FORC features discussed in the text.

\* Electronic address: wuyizheng@fudan.edu.cn